\documentstyle[manuscript,aps]{revtex}
\input epsf

\newcommand{\beq}{\begin{equation}}
\newcommand{\eeq}[1]{\label{#1} \end{equation}}
\newcommand{\beqar}{\begin{eqnarray}}
\newcommand{\eeqar}[1]{\label{#1} \end{eqnarray}}

\def\beq{\begin{equation}}
\def\eeq{\end{equation}}
\def\bea{\begin{eqnarray}}
\def\eea{\end{eqnarray}}
\def\R{{\bf R}}
\def\Z{{\bf Z}}
\def\RP{{\bf RP}}
\def\S{{\bf S}}
\begin{document}
\draft
%\preprint{SNUTP-98-088, SOGANG-HEP 242/98, APCTP-98-020, hep-th/9808182}
\setcounter{page}{0}
\title{\Large\bf $SO(2N)$ $(0, 2)$ SCFT and M Theory on 
$AdS_7 \times {\bf RP}^4$ } 
\author{\large \rm 
Changhyun Ahn\footnote{chahn@spin.snu.ac.kr}}
\address{ \it Center for  Theoretical Physics, 
Seoul National University, 
Seoul 151-742, Korea} 
\author{\large \rm
Hoil Kim\footnote{ hikim@gauss.kyungpook.ac.kr}} 
\address{\it Topology and Geometry Research Center, 
Kyungpook National University, 
Taegu 702-701, Korea}
\author{\large  Hyun Seok Yang\footnote{hsyang@physics4.sogang.ac.kr}}
\address{ \it Department of Physics, 
Sogang University, 
Seoul 121-742, Korea}
\date{\today}

\maketitle
\begin{abstract}
We study M theory on $AdS_7 \times \RP^4$ corresponding to 6 dimensional 
$SO(2N)$
$(0, 2)$ superconformal field theory on a circle which becomes 
5 dimensional super Yang-Mills theory 
at low energies. For $SU(N)$ $(0, 2)$ theory, a wrapped D4 brane 
on $\S^4$ which is connected to a D4 brane on the boundary 
of $AdS_7$ by  $N$ fundamental strings can be 
interpreted as baryon vertex. 
For $SO(2N)$ $(0, 2)$ theory, by using the property of 
homology of $\RP^4$, 
we classify various wrapping branes.
Then we consider particles, strings, twobranes, domain walls and the baryon 
vertex in Type IIA string theory. 
\end{abstract}

\setcounter{footnote}{0}

%*****************************************************
%*****************************************************
%*****************************************************

\section{Introduction}
\setcounter{equation}{0}

In \cite{mal}  the large $N$ limit of superconformal field theories (SCFT) 
was described by taking the
supergravity limit on anti-de Sitter (AdS) space.
One can obtain the scaling dimensions of operators of SCFT from the 
masses of particles in string/M theory \cite{polyakov}. 
In particular, 
${\cal N}=4$ super $SU(N)$ Yang-Mills theory in 4 dimensions is described by
Type IIB string theory on $AdS_5 \times {\bf S}^5$. 
This can be done by considering the brane configuration in the string 
theory and taking
a low energy limit which decouples the field theory from gravity and 
simultaneously
considering the near horizon geometry of the correponding supergravity 
solution.
There are ${\cal N} =2, 1, 0$ superconformal theories in 
4 dimensions which have corresponding supergravity description 
on orbifolds of $AdS_5 \times {\bf S}^5$
\cite{kachru}. This proposed duality was tested by studying
the Kaluza-Klein (KK) states of supergravity theory and 
by comparing them with the chiral primary operators
of the SCFT on the boundary \cite{ot} and used to calculate the energy of
quark-antiquark pair \cite{ryma}.

The field theory/M theory duality provides
a supergravity description on $AdS_4$ or $AdS_7$ 
for some superconformal theories
in 3 and 6 dimensions, respectively. The maximally supersymmetric theories
have been studied in \cite{aoy,lr} and
the lower supersymmetric case was also realized on the worldvolume of 
M theory at orbifold singularities \cite{fkpz}.
The KK spectrum description on the twisted states of $AdS_5$ orbifolds
was discussed in \cite{gukov}. 

Recently, it was observed that
the gauge group is replaced by $SO(N)/Sp(N)$ \cite{witten1} by taking
appropriate orientifold operations for the string theory on $AdS_5 \times \RP^5
$ (See also \cite{aoy,kaku}). 
By analyzing the discrete torsion for Neveu Schwarz $B$ field and RR $B$
field, many features of gauge theory were described by various wrapping branes.

A system of $N$ D4 branes in the decoupling limit 
can be well described by considering 
$N$ M5 branes wrapped on the eleventh dimensional circle in M theory by taking
the limit of eleven dimensional Planck length being zero and keeping the 
radius of circle fixed \cite{imsy}. See also
relevant papers \cite{bst,abks}. Then $(0, 2)$ six dimensional CFT on a 
circle becomes five dimensional super Yang-Mills theory at low energy.
The UV region is described by the (0,2) SCFT on a circle which is dual for 
large $N$ to M theory on a background  $AdS_7 \times \S^4$. 
In the IR region, the theory is described by five dimensional super Yang-Mills theory  having a dual supergravity description by the Type IIA D4 brane 
solution.
 
In this paper, we generalize the work of \cite{witten1} to the
case of $AdS_7 \times \RP^4$ where the eleventh dimensional circle is one of
$AdS_7$ coordinates. So we are dealing with $(0, 2)$ six dimensional SCFT
on a circle rather than uncompactified full M theory. 
In section II,
for $SU(N)$ $(0, 2)$ theory, a wrapped D4 brane on $\S^4$ can be 
interpreted as baryon vertex.  There exist $N$ fundamental strings 
connecting a D4 brane on the boundary 
of $AdS_7$ with the D4 brane on $\S^4$. By putting $N$ M5 branes in the 
$\R^5/\Z_2$ orbifold singularity, where the ${\Z}_2$ acts by a
reflection of the 5 directions transverse to the M5 branes, and also
by changing the sign of the 3-form field $C_3$, we will obtain the large
$N$ limit of the $SO(2N)$ (0,2) SCFT and $\RP^4$ orientifold after
removing the $\R^5/\Z_2$ orbifold singularity. 
Then using the property of homology of $\RP^4$ 
we classify various wrapping branes and discuss their topological
restrictions.
In section III,
we consider particles, strings, twobranes, domain walls and the baryon 
vertex in Type IIA string theory. 
Finally, in section IV, we will come to the summary of this paper and
comment on the future direction.

%*****************************************************
%*****************************************************
%*****************************************************

\section{ $SO(2N)$ $(0, 2)$ SCFT and Branes on ${\bf RP^4}$  }
\setcounter{equation}{0}

\subsection{The Baryon Vertex in $SU(N)$}

Let us consider M theory on $\R^{5} \times \R^5 \times \S^1 $. For first
$\R^5$, we can take $(x^0, x^1, x^2, x^3, x^6)$ directions and for second
$\R^5$, we take $(x^4, x^5, x^7, x^8, x^9)$ transverse to M5 branes. 
The eleventh coordinate
$x^{10}$ is compactified on a circle $\S^1$ and is 
a periodic coordinate of period $2 \pi$. 
For small radius of $\S^1$, we can regard M theory as Type IIA
string theory which can be 
described in the context of $AdS_7 \times \S^4$ where the D4
branes are M5 branes wrapping on $\S^1$. The radial function
$\rho= \sqrt{ (x^4)^2 +(x^5)^2+(x^7)^2+(x^8)^2+(x^9)^2} $ of second $\R^5$
will be one of the $AdS_7$ coordinates, the other six being the ones in 
$\R^5 \times \S^1$.
The $AdS_7 \times \S^4$ compactification has $N$ units of
four-form flux on $\S^4$ as follows.
\bea
\int_{\S^4} \frac{G_4}{2 \pi}=N,
\label{n}
\eea 
where $G_4$ is four-form field. When a D4 brane is wrapped on $\S^4$, 
on the worldvolume of the D4 brane, 
there exists a $U(1)$ gauge field which can 
couple $G_4$ by 
\bea
\int_{\S^4 \times {\bf R}} a \wedge \frac{G_4}{2 \pi},
\label{couple}
\eea
where ${\bf R}$ is a timelike curve in $AdS_7$.
From these two relations (\ref{n}) and (\ref{couple}), 
$G_4$ field has the $N$ units of $a$ charge.
There should be $-N$ units of $a$ charge somewhere in order to 
satisfy the vanishing of $a$ charge in a closed universe.
This can be done by putting oriented $N$ fundamental strings ending on
the D4 brane. 
So we can interprete the wrapped D4 brane as baryon vertex or 
antibaryon vertex\footnote{There are another arguments supporting 
this baryon picture that
seven dimensional Chern-Simons term on $AdS_7$ produces a M2 brane with
$N$ units of charge and, wrapping on the eleventh circle $\S^1$, 
gives $N$ open strings joining together 
in the $AdS_7$ space \cite{grossooguri} 
and, in \cite{abks}, the baryon
vertex was constructed in full M theory by the M5 brane wrapped on
$S^4$ balanced by $N$ M theory membranes ending on it. Our picture
corresponds to the compactification along the eleventh circle of theirs.}. 
The energy of baryon was calculated as a function of its size in
\cite{bisy} along the similar picture.

In $SU(N)$ gauge theory, the quantity of gauge invariant combination
of $N$ quarks should be completely antisymmetric.
This antisymmetry of baryon vertex corresponding to the behavior of
fundamental string connecting D4 brane to the boundary of $AdS_7$ space can be
understood with the following boundary conditions. 
The time zero section of $AdS_7 \times
\S^4$ is a copy of $\R^5 \times \S^1 \times \S^4$. 
We can consider a static D4 brane
which has the worldvolume of ${\widetilde \S}^4 \times R$ 
where $R$ is a point of $\S^4$
and ${\widetilde \S}^4$ is a large four sphere near infinity 
in $\R^5 \times \S^1$. 
Then our $N$ strings
connnect between D4 brane on ${\widetilde \S}^4 \times R$ with 
D4 brane on $Q \times \S^4$ where $Q$ is a point in a time zero slice
of $AdS_7$. 
The former D4 brane and the latter one are
linked in the $\R^5 \times \S^1 \times \S^4$, but they do not
intersect each other. 
They have linking number $\pm 1$.
The string stretching between linked D branes 
has eight mixed Dirichlet-Neumann boundary conditions.
Then the string zero point energy
is 0 in the Ramond sector as always, and 
the zero point energy in NS sector is positive, 1/2. 
Thus true ground state is a Ramond state, whose Fock space generates
Clifford algebra.
Thus the strings stretching between the boundary (or a D4 brane)
and the D4 brane are certainly fermionic strings. 

\subsection{The ${\bf RP^4}$ Orientifold}

It was observed \cite{aoy}
that the large $N$ limit of $SO(2N)$ $(0, 2)$ SCFT in 6 
dimensions
corresponds to the low energy theories of $N$ M5 branes coinciding
at $\R^5/\Z_2$ orientifold singularity \cite{witten3}.
Let us consider M theory on $\R^5  \times \R^5/\Z_2 \times \S^1$.  
The $\Z_2$ acts by sign change on 
all five coordinates in second $\R^5$ as follows:
$
(x^4, x^5, x^7, x^8, x^9) \rightarrow 
(-x^4, -x^5, -x^7, -x^8, -x^9).
$ 
The angular directions in $\R^5/\Z_2$ are identified with
$\RP^4 $.
There are $N$ parallel M5 branes which are sitting 
at an orientifold four-plane (O4 plane) 
which is located at $x^4=x^5=x^7=x^8=x^9=0$. 
For small radius of $\S^1$, we can regard this as Type IIA
orientifold on $\R^5/\Z_2$ which can be 
described in the context of $AdS_7 \times \S^4$. 
The orientifolding operation replaces four sphere $\S^4$
around the origin in $\R^5$ with $\RP^4=\S^4/\Z_2$. We will
describe how $SO(2N)$ $(0, 2)$ SCFT in 6 dimensions on a circle
can be interpreted as M theory on $AdS_7 \times \RP^4$ where the eleventh 
dimensional circle is in $AdS_7$ space.  
This is our main goal in this paper.

Let us study the property of  $AdS_7 \times \RP^4$ orientifold.
Let $x$ be the generator of $H^1(\RP^4, \Z_2)$ which is isomorphic
to $\Z_2$,\footnote{ Notice that  $H_p(\RP^n, \Z_2)=\Z_2, 0 \leq p \leq n$
and $H^p(\RP^n, \Z_2)=\Z_2, 0 \leq p \leq n$.} 
$\Sigma$ be a string
worldsheet and $w_1(\Sigma) \in H^1(\Sigma, \Z_2)$ be the obstruction
to its orientability. Then we only consider the map $\Phi: \Sigma \rightarrow
AdS_7 \times \RP^4$ such that $\Phi^*(x)=w_1(\Sigma)$.
Since $\Z_2$ action on $\S^4$ is free (no orientifold fixed points),
there is no open string sector. 
An unorientable closed string worldsheet $\Sigma=\RP^2$ can be 
identified with the quotient of the two sphere $\S^2$
by the overall sign change.
The map $\Phi: \RP^2 \rightarrow \RP^4$ satisfying the constraints
$\Phi^*(x)=w_1(\RP^2)$ is the embedding 
$(x_1, x_2, x_3) \rightarrow (x_1, x_2, x_3, 0, 0)
$.

In M theory, M2 brane and M5 brane are electrically and
magnetically charged objects with respect to the three-form potential
$C_3$. The Dirac quantization condition for these objects takes the
form of flux quantization condition \cite{witten2}
for the four-form field strength
$G_4=dC_3$:
\bea
2\int_D \frac{G_4}{2\pi} \equiv 
\int_D w_4 \;\; \mbox{mod} \;\; 2,
\label{fluxQ}
\eea
where $D$ is a four-cycle, a copy of ${\RP^4}$, and 
$w_4$ is the fourth Stieffel-Whitney class
of the eleven-dimensional space-time. 
This has a direct consequence in M theory on ${\bf R}^5/{\bf Z}_2$
orbifold, which implies that the ${\bf Z}_2$ fixed plane itself
carries the M5 brane charge $-1$. Moreover, one cannot
put odd number of M5 branes on the top of it.

It was shown \cite{witten1} 
that there exist four kinds of models in Type IIB string theory
classified by the discrete torsion of the two two-form fields,
NS $B$ field $B_{NS}$ and RR $B$ field $B_{RR}$. The results are
summarized as follows:
\bea
 i) & \; SO(2N)\; \mbox{gauge group}: & (\theta_{NS}, \theta_{RR})=(0,0), 
\nonumber\\
ii) & \; SO(2N+1)\; \mbox{gauge group}: & (\theta_{NS},
\theta_{RR})=(0, 1/2), \nonumber\\
iii) & \; Sp(2N)\; \mbox{gauge group}: & (\theta_{NS}, 
\theta_{RR})=(1/2,0), \nonumber\\  
iv) & \; Sp(2N)\; \mbox{gauge group}: & (\theta_{NS},
\theta_{RR})=(1/2,1/2),
\eea
where $(\theta_{NS}, \theta_{RR})$ are two types of discrete
torsion for $B_{NS}$ and $B_{RR}$ respectively.

It is easy to see that 
the Type IIA brane configurations and its M theory realization considered in 
\cite{hori} can be obtained from the type IIB configurations 
by Witten \cite{witten1} by taking T-duality along the $x^6$ direction. 
The M theory realization of the type IIA O4 planes comes from the two
kinds of orientifold, ${\bf R}^5 \times ({\bf R}^5 
\times {\bf S}^1)/{\bf Z}_2$ and ${\bf R}^5 \times {\bf R}^5/{\bf Z}_2
\times {\bf S}^1$ where $\Z_2$ action on the eleventh circle ${\bf S}^1$ is
$x^{10} \rightarrow x^{10} +\pi$. 
In the first case, since the ${\bf Z}_2$ action is
free, there is no orientifold fixed point which so generates no
M5 brane charge, while in the second case, the orientifold has a singular
fixed plane and so carries M5 brane charge $-1$ originated by 
the singularity. 
In particular, we should have a single D4
brane stuck on the $O4^{-}$ plane to get $SO(2N+1)$ gauge group.
In order to obtain the O4 planes with $0$ or $+1$ D4
brane charge, we should stick one or two M5 branes on the
orientifold in the way consistent with the flux quantization 
condition (\ref{fluxQ}). Then we can identify
the consistent M theory realization of four types O4 planes 
obtained by Hori \cite{hori}:
\bea
i) & \; SO(2N)\; \mbox{with} \; O4^-: \;
& \mbox{M theory on ${\bf R}^5 \times {\bf R}^5/{\bf Z}_2 
\times {\bf S}^1$}\nonumber\\
ii) & \; SO(2N+1)\; \mbox{ with } \; O4^0: \; 
& \mbox{M theory on ${\bf R}^5 \times 
({\bf R}^5 \times {\bf S}^1)/{\bf Z}_2$}\nonumber\\
iii) & \; Sp(2N)\; \mbox{ with}\; O4^+: \; 
& \mbox{M theory on ${\bf R}^5 \times 
{\bf R}^5/{\bf Z}_2 \times {\bf S}^1$} \nonumber\\  
iv) & \; Sp(2N)\; \mbox{ with} \; \widetilde{O4}^+: \;
& \mbox{M theory on}\;
\mbox{${\bf R}^5 \times ({\bf R}^5 \times {\bf S}^1)/{\bf Z}_2$ }.  
\label{four}
\eea   
There exist a pair of M5 branes at the $\Z_2$ fixed plane for case $iii)$
and a single M5 brane is stuck on the $\Z_2$ invariant cylinder 
for case $iv)$.

We will demonstrate that the
M theory realization of the four kinds of model in Type IIA string theory can
be obtained from the topology of $B_{NS}$ and the RR $U(1)$
Wilson line, by using that the relation of the $B_{NS}$ field 
and the three-form potential $C_3$ (See (\ref{gh}))
and by considering the complex
line bundle associated with the RR one-form $A_{RR}$ defined on the circle 
${\bf S}^1$. 
Since the $SO(N)/Sp(N)$ gauge theories for large $N$ 
can be distinguished by the
sign of the string ${\bf RP}^2$ diagram,
this can be classified by the topology of the $B_{NS}$ field.
When we consider M theory
on $(\R^5 \times {\bf S}^1)/ {\bf Z}_2$ or 
$\R^5/ {\bf Z}_2 \times {\bf S}^1$, the ${\bf Z}_2$ involution 
possibly induces a non-trivial holonomy (RR $U(1)$ Wilson line) for the
former case which corresponds to Type IIA string
theory on $\R^5/{\bf Z}_2$, this can be classified by the topology
of the $A_{RR}$ field. 
Since the ${\bf Z}_2$ action flips the sign of $G_4=dC_3$ and 
preserves the orientation of the circle ${\bf S}^1$, 
the ${\bf Z}_2$ action in ten dimensions flips the sign of the 
$H_{NS}=dB_{NS}$ according to the relation
\bea
\int_{{\bf S}^1} \frac{G_4}{2\pi}= \frac{H_{NS}}{2\pi}.
\label{gh}
\eea
This means that a cohomology class $[H_{NS}]$ takes values in a
twisted integer coefficient ${\widetilde {\bf Z}}$ where the twisting is
determined by an orientation bundle. Since the RR $U(1)$
gauge field $A_{RR}$ represents a connection of the $U(1)$ bundle 
over $\R^5/{\bf Z}_2$ where
$U(1)$ group is identified with the circle ${\bf S}^1$, 
the holonomy of this complex line bundle can be measured by the field
strength $F_{RR}=dA_{RR}$. Then the relevant cohomology
groups\footnote{ Note that there is
an isomorphism pairing $p$-th homology group $H_p(M, R)$ and $(n-p)$-th
cohomology group $H^{n-p}(M, R)$ for all $p$ and modulus $R$ iff $M$ is an
orientable compact manifold, which is known as Poincar\'e duality:
$H_p(M, \Z) \approx H^{n-p}(M, \Z)$ and $H_p(M, \widetilde{\Z}) 
\approx H^{n-p}(M, \widetilde{\Z})$. If $H_n(M, \Z)=\Z$, the $n$-dimensional
manifold $M$ is orientable. For an odd
dimensional projective space $\RP^n$,  the Poincar\'e
duality works since it is orientable.}
measuring the topological types of the fields 
$B_{NS}$ and $A_{RR}$ are given by respectively
\bea
H^3({\bf RP}^4, {\widetilde {\bf Z}})\approx {\bf Z}_2,\qquad
H^2({\bf RP}^4, {\bf Z})\approx {\bf Z}_2.
\eea 
If we denote the values of the cohomologies 
$H^3({\bf RP}^4, {\widetilde {\bf Z}})$ and $H^2({\bf RP}^4, {\bf Z})$ as
$(\alpha, \beta)$ respectively, 
we get the topological classification of the four
models:
\begin{eqnarray}
\begin{array}{ll} 
O4^- :(\alpha, \beta)=(0,0),\qquad & O4^0 :(\alpha, \beta)=(0,1),  \\
O4^+ :(\alpha, \beta)=(1,0),\qquad & \widetilde{O4}^+ :(\alpha, \beta)=(1,1).
\end{array} 
\end{eqnarray}

Then how can we obtain full uncompactified M theory on $AdS_7 \times
\RP^4$ from Type IIA string theory,
which means we take the limit of eleventh circle radius being infinite?
There is a big difference between the case $i)$ and remaining
$ii), iii)$ and $iv)$ in (\ref{four}).
The $O4^{-}$ plane remains unchanged under the T-S-T transformation where 
T-duality is taken along the $x^9$ direction \cite{gimon}. 
The $O4^{0}$ and $\widetilde{O4}^{+}$  
do not produce orientifold plane under T-S-T chain.
The $O4^{+}$ becomes $O4^{0}$ with doubly wrapped D4 brane 
under the T-S-T transformation which implies
under the Montonen-Olive duality they are exchanged each other.
We have seen there are two classes of Type IIA string theory corresponding to
our M theory configuration which has AdS/CFT correspondence: $i)$ and $iii)$.
The difference between these is whether two M5 brane are stuck on the $\Z_2$
fixed plane or not. This M5 brane pair may be separated from the $\Z_2$
fixed plane without cost of energy, if the radius of circle becomes very 
large (uncompactified M theory). Then there will be no difference 
between $i)$ and $iii)$ as long as the AdS/CFT 
correspondence is valid and thus the full uncompactified M theory has
only one type of orientifold 5-plane as speculated in \cite{gimon}.

\subsection{Various Wrapped Branes}

Now we consider the possibilities of brane wrapping on ${\bf RP}^4$ 
in the Type IIA string theory. 
We first recall that fundamental strings, D4 branes and D6 branes 
in Type IIA string theory are M2 branes,
M5 branes and KK monopoles 
wrapped around the circle ${\bf S}^1$ in M theory respectively.
But our dimension counting of branes in $AdS_7$ is in viewpoint of
Type IIA string theory, we hope it does not cause any confusion. 

The wrappings of string and NS5 brane are 
classified by the twisted
homology $H_i({\bf RP}^4, {\widetilde {\bf Z}})$ for wrapped on an
$i$-cycle in ${\bf RP}^4$ along the line of \cite{witten1}:\\
$(i)$ unwrapped string, giving a onebrane in $AdS_7$.\\
The wrapping modes that would give zerobranes are not possible
since $H_1({\bf RP}^4, {\widetilde {\bf Z}})=0$.\\
$(ii)$ unwrapped NS5 brane remains a fivebrane in $AdS_7$,\\ 
$(iii)$ wrapped on a two-cycle, to give a threebrane 
in $AdS_7$, classified by $H_2({\bf RP}^4, {\widetilde {\bf Z}})={\bf Z}_2$,\\
$(iv)$ wrapped on a four-cycle, to give a onebrane 
in $AdS_7$, classified by $H_4({\bf RP}^4, {\widetilde {\bf Z}})={\bf Z}_2$.
The wrappings of D2 and D4 brane are also classified by the twisted
homology $H_i({\bf RP}^4, {\widetilde {\bf Z}})$ for wrapped on an
$i$-cycle in ${\bf RP}^4$ since the twobrane charge is odd under the
orientifolding operation (it comes from the fact that, in M theory, 
${\bf Z}_2$ action flips the sign of three-form $C_3$) and the
fourbrane is dual to the twobrane:\\
$(v)$ unwrapped D2 brane, giving a twobrane in $AdS_7$,\\
$(vi)$ wrapped on a two-cycle, to give a zerobrane 
in $AdS_7$, classified by $H_2({\bf RP}^4, {\widetilde {\bf Z}})={\bf Z}_2$,\\ 
$(vii)$ unwrapped D4 brane remains a fourbrane in $AdS_7$,\\
$(viii)$ wrapped on a two-cycle, to give a twobrane 
in $AdS_7$, classified by $H_2({\bf RP}^4, {\widetilde {\bf Z}})={\bf Z}_2$,\\
$(ix)$ wrapped on a four-cycle, to give a zerobrane 
in $AdS_7$, classified by $H_4({\bf RP}^4, {\widetilde {\bf Z}})={\bf Z}_2$. 

The $k$ units of KK momentum mode around the 
eleventh circle ${\bf S}^1$ can be identified as $k$ D0 branes which
is charged BPS particles. In the five dimensional gauge theory
context, they appear as charge $k$ instantons on the D4 brane 
worldvolume \cite{seiberg}. 
We cannot consider wrapping modes on ${\bf RP}^4$ of D0 branes, since
the eleventh circle ${\bf S}^1$ is extended in $AdS_7$, 
not ${\bf RP}^4$. Similarly, we cannot consider a completely unwrapped
sixbrane on ${\bf RP}^4$, since we consider the D6 brane as the
KK monopole in M theory which is magnetically charged under
the gauge field $A_\mu=G_{\mu 10}$. The wrappings of D6 brane 
are classified by the
ordinary (untwisted) homology $H_i({\bf RP}^4, {\bf Z})$ for wrapped on an
$i$-cycle in ${\bf RP}^4$ since the sixbrane is dual 
to the zerobrane:\\
$(x)$ wrapped on a one-cycle, to give a fivebrane 
in $AdS_7$, classified by $H_1({\bf RP}^4, {\bf Z})={\bf Z}_2$,\\
$(xi)$ wrapped on a three-cycle, to give a threebrane 
in $AdS_7$, classified by $H_3({\bf RP}^4, {\bf Z})={\bf Z}_2$.

According to the similar argument done in Type IIB description \cite{witten1}, 
one can derive a 
topological restriction on the brane wrappings on ${\bf RP}^4$ 
just described. In particular, the topological restriction coming from
the holonomy of the connection $A_{RR}$ on the complex line bundle is
trivial since the two cases, $O4^-$ and $O4^+$, we are interested in
M theory context have the trivial RR $U(1)$ Wilson line. 
Nevertheless, we will show
there are interesting possibilities that, in intersection of NS5
branes or D6 branes with the O4 plane, the type of the orientifold
plane is changed. 
Therefore it is sufficient only to consider the discrete torsion 
$\theta_{NS}$ of the field $B_{NS}$. We will show that 
the description of brane wrappings on ${\bf RP}^4$ 
is consistent with the topological
restriction coming from the RR discrete torsion $\theta_{RR}$, which
should vanish in our case.

In the case $(iii)$, there is no restriction on wrapping of NS5
branes on ${\bf RP}^2 \subset {\bf RP}^4$, to make a threebrane in
$AdS_7$, since $H^2({\bf RP}^2, {\widetilde {\bf Z}})={\bf Z}$.
In the case $(iv)$, the NS5 brane can be wrapped on 
${\bf RP}^4$, to make a string in $AdS_7$, only if $\theta_{RR}=0$, 
since $H^2({\bf RP}^4, {\widetilde {\bf Z}})=0$.
Similarly, in the case $(vi)$ and $(viii)$, there is no restriction
on wrapping of D2 and D4 branes on ${\bf RP}^2 \subset {\bf RP}^4$, 
to make a particle and a twobrane in $AdS_7$ respectively,
since $H^2({\bf RP}^2, {\widetilde {\bf Z}})={\bf Z}$. 
In the case $(ix)$, the D4 brane can be wrapped on 
${\bf RP}^4$, to make a particle in $AdS_7$, only if $\theta_{NS}=0$, 
since $H^2({\bf RP}^4, {\widetilde {\bf Z}})=0$.
Similarly, in the case $(xi)$, the D6 brane can be wrapped on 
${\bf RP}^3 \subset {\bf RP}^4$, to make a threebrane in $AdS_7$, 
only if $\theta_{NS}=0$, since $H^2({\bf RP}^3, {\widetilde {\bf Z}})=0$.
In the case $(x)$, there is no restriction on wrapping of D6 
branes on ${\bf RP}^1$, to make a fivebrane in $AdS_7$, 
since in this case ${\bf RP}^2$ cannot be
deformed in the D6 brane.

In the cases $(iii)$ and $(iv)$, the system of an O4 plane
intersecting with a single NS5 brane is possible. We should remember
that the NS5 brane has a unit magnetic charge under the NS two-form
field $B_{NS}$, but does not carry electric or magnetic charge for the
RR one-form $A_{RR}$. Thus, in crossing such a fivebrane, the value of
$\theta_{NS}$ jumps by one unit, this means that the type of the
orientifold is changed as shown in \cite{evan-elit}.
The possible configurations are 
$O4^--NS5-O4^+$ and $O4^0-NS5-\widetilde{O4}^+$ by charge
conservation. The similar argument can be applied in the system of an
O4 plane intersecting with a single D6 brane. This comes from 
the fact that the D6 brane is a magnetic monopole for the RR $U(1)$
gauge field $A_{RR}$. The case $(xi)$ exactly corresponds to the same 
configuration considered in \cite{hori}, where it was shown that 
only $O4^--D6-O4^0$ and
$O4^+-D6-\widetilde{O4}^+$ configurations are the allowed patterns of
dividing an O4 plane by a single D6 brane. This O4-D6 system can be
understood by considering a Taub-NUT space realizing this
configuration and $\Z_2$ orbifolding of M theory on this space. 
Since the case $(x)$ corresponds to the situation the O4 plane is
completely embedded in the D6 worldvolume which fills out the full
$AdS_7$ space, there is no difference of magnetic RR $U(1)$ charge 
around the O4 plane. Therefore, this case does not cause any change
in the type of O4 planes.
 
%**********************************************************
%**********************************************************
%**********************************************************

\section{ Gauge Theory and Branes on $\RP^4$}
\setcounter{equation}{0}

\subsection{The Pfaffian Particles}

Let us consider particles coming from wrapped D2 brane on $\RP^2 \subset
\RP^4$ \cite{abks} 
classified by $H_2(\RP^4, \widetilde{\Z})=\Z_2$. 
We are considering the case of $vi)$ in section IIC. 
The number of particles is conserved modulo two and a pair of such particles 
can annihilate.
Since a D2 brane can wrap on $\RP^2$ without any topological restriction for
$\theta_{NS}$, 
we should find a corresponding operator for each gauge group 
$SO(2k)$ and $Sp(k)$, where $k$ is the number of 
four-form flux quanta on $\RP^4$: 
\bea
\int_{\RP^4} \frac{G_4}{2\pi}=k,
\eea
where it is understood that on the double cover $\S^4$,
the number of quanta is $N=2k$.
We expect along the similar line suggested in \cite{witten1} 
that the D2 brane wrapped on $\RP^2$ should be identified
with the Pfaffian particle of $SO(2k)$ theory where Pf$(\Phi)=\epsilon^{
a_1 a_2 \cdots a_{2k}} \Phi_{a_1 a_2} \cdots \Phi_{a_{2k-1} a_{2k}}$ is an 
irreducible gauge invariant polynomials of order $k$ and $\Phi_{ab}, a,b=
1, 2, \cdots, 2k$ is an antisymmetric second rank tensor in the adjoint 
representation of $SO(2k)$.

In order to identify this with Pfaffians, we will determine the quantum
numbers of the low lying states on $AdS_7$ and SCFT sides 
under R symmetry group of the theory. 
The manifold $\RP^4$ has a symmetry group $G_0=SO(5)/\Z_2$.
A subspace $\RP^2$ is invariant under $H_0=SO(3)\times SO(2)/\Z_2$. The space
of this embedding is the homogeneous space $G_0/H_0$ that is the same
as $G/H$ where $G=SO(5)$ and $H=SO(3) \times SO(2)$.
The quantum states are not ordinary functions on $G/H$ but sections of 
a line bundle of degree $k$.
A section of a line bundle on $G/H$ is a function on the $G$
manifold. 
When we identify $G=SO(5)$ as the group of $5 \times 5$ orthogonal
matrices $g^i_j$ where $i, j=1, 2, \cdots,  5$, 
one can expand functions on the
$G$ manifold as polynomials in the matrix elements of $G$. The matrix
element of $g^i_j$ transforms as $({\bf 5}, {\bf 5})$ under 
$SO(5)\times SO(5)$ and as $({\bf 5}, {\bf 3})_{{\bf 0}} 
\oplus ({\bf 5}, {\bf 1})_{{\bf 1}}
\oplus ({\bf 5}, {\bf 1})_{{\bf -1}} $ under $SO(5)\times SO(3)\times SO(2)$.
We denote $SO(2)$ charge in the subscript. The polynomials of degree $k$
in the $({\bf 5}, {\bf 1})_{{\bf 1}}$ transform  in the traceless 
symmetric product of $k$ copies of ${\bf 5}$.
These are holomorphic sections of the line bundle.
On the other hand, in the boundary SCFT, the chiral operators
from the scalar fields can be constructed. The scalars are transforming
in the ${\bf 5}$ of the $SO(5)$ global symmetry and in the adjoint 
representation of group $SO(2k)$.
The Pfaffian of the scalars, 
 Pf$(\Phi)=\epsilon^{
a_1 a_2 \cdots a_{2k}} \Phi^{i_1}_{a_1 a_2} \cdots 
\Phi^{i_k}_{a_{2k-1} a_{2k}}$
where the $i$'s  are $SO(5)$ indices, transforms in the 
$k$-th symmetric product of the ${\bf 5}$.

What is an appropriate operator of an $Sp(k)$ gauge theory 
corresponding to the D2 brane wrapped on
an ${\RP}^2$ subspace of ${\RP}^4$. This operator will be quite
different from the Pfaffian operator in $SO(2k)$ theory
since there is no such operator in $Sp(k)$ theory.
We speculate that the candidate may be given by the holomorphic gauge
invariant polynomial\footnote{In four dimensional $Sp(k)$ gauge
theory \cite{lest}, this appears as an element of chiral 
ring operators under an superpotential by an adjoint matter.} 
in an adjoint field
$\Phi_{ab}=\Phi_{ba},\;a,b=1,\cdots,2k$, that is, 
$U_k(\Phi)=\frac{1}{k} \mbox{Tr}\Phi^k$, where
trace is taken with the invariant second rank antisymmetric tensor 
$\gamma^{ab}$. Note that the operator $U_k(\Phi)$ vanishes for odd $k$. 
The quantum numbers of the state
can be determined by the same method as $SO(2k)$ gauge theory. 
But, we do not find any obvious reason for the modulo two conservation
of the number of such quanta.

\subsection{Strings and Twobranes}

Let us consider ``solitonic'' strings in $AdS_7$ coming 
from wrapped NS5 brane on
$\RP^4$ which is the case of $iv)$. 
According to the topological restriction,
$\theta_{RR}=0$, the solitonic string is possible for both 
orthogonal and symplectic group and can induce the sign flip of an
orientifold plane, intersecting with the plane. 
The solitonic strings can annihilate in pairs due to 
$H_4({\bf RP}^4, {\widetilde {\bf Z}})={\bf Z}_2$.
Let the NS5 brane worldvolume coordinates be 
$(x^0, x^4, x^5, x^7, x^8, x^9)$ directions and
be specified by $x^1=x^2=x^3=x^6=0$. 
Then the solitonic string worldsheet in $AdS_7$ is parametrized by the radial 
function $\rho$ of $\RP^4$ and $x^0$ and is at $x^1=x^2=x^3=x^6=0$.    
The tension of this string is proportional to the NS5 brane tension, 
of order $1/\lambda^2$ and stretched to infinity in the radial
direction of $AdS_7$. It is not obvious that the solitonic string can
carry an external spinor charge as the fat string in $AdS_5$ 
\cite{witten1} and it is not clear how to interpret this 
in the boundary SCFT.  

Let us consider D4 brane whose worldvolume is specified 
by $x^1=x^2=x^3=x^8=x^9=0$,
with arbitrary values of $x^0$ and of $x^4, x^5, x^6, x^7$. 
This is the case of $viii)$.
From the $AdS_7 \times
\RP^4$ point of view such a D4 brane is wrapped on an 
$\RP^2 \subset \RP^4$ and
looks like twobrane in $AdS_7$. The $\RP^2$ is the subspace of 
$\RP^4$ with
$x^8=x^9=0$. The twobrane worldvolume in $AdS_7$ is parametrized 
by  radial
function $\rho$
of $\R^5/\Z_2$, $x^6$ and $x^0$ and is at $x^1=x^2=x^3=x^8=x^9=0$.
The 4-4 strings connecting the D4 brane in $AdS_7$ 
($(x^0, x^1, x^2, x^3, x^6)$ directions) 
to D4 brane in $(x^0, x^4, x^5, x^6, x^7)$ directions are 
fermionic strings.
Since D4 brane wrapped on $\RP^2 \subset \RP^4$ and D4 brane 
in $AdS_7$ meet
at $x^1=x^2=\cdots=x^5=x^7=x^8=x^9=0$, 
the ground state of the 4-4 string has 
zero energy. The ground
states of these strings give $N$ fermionic zero modes in 
the spinor representation of 
$SO(N)$. The D4 brane regarded as a twobrane in $AdS_7$ 
has an end point at $\rho=0$.
This endpoint lies on the boundary of $AdS_7$ at which there are
external spinor charges. 
At large distances, this twobrane may be used to define ``Wilson
surface'' observable with external spinor charge 
of the low energy $(0,2)$ SCFT.

\subsection{Domain Walls}

Let us consider the objects in $AdS_7\times {\bf S}^4$ and 
$AdS_7\times {\bf RP}^4$ that look like fivebranes in the seven
noncompact dimensions of $AdS_7$. Since the $AdS_7$ has six spatial
dimensions, the fivebrane could potentially behave as a domain wall,
with the string theory vacuum ``jumping'' as one crosses the
fivebrane. In $AdS_7\times {\bf S}^4$ and $AdS_7\times {\bf RP}^4$, 
the only such object is the Type IIA NS5 or D4 branes, which  come 
from the M5 brane. 
Contrary to Type IIB theory, we cannot obtain the domain wall made by
wrapping a brane on $\RP^i \subset {\bf RP}^4$ as shown in section IIC. 

Since the fivebrane is the magnetic source of the four-form field $G_4$ 
in M theory, the integrated four-form flux over ${\bf S}^4$ 
or ${\bf RP}^4$ jumps by one unit when one
crosses the fivebrane. This means that the gauge group of the boundary
conformal field theory can change, for example, from $SU(N)$ on one side to 
$SU(N\pm 1)$ on the other side for $AdS_7\times {\bf S}^4$. 
While, for $AdS_7\times {\bf RP}^4$, the change of the gauge group
depends on which brane one considers. 
If one is crossing the D4 brane, it changes from $SO(N)$ to $SO(N\pm 2)$ 
or from $Sp(N/2)$ to $Sp(N/2\pm 1)$ since the D4 brane charge changes 
by two units on double cover. While, if one is crossing the NS5 brane
along the eleventh circle ${\bf S}^1$, there is a jump in discrete 
torsion $\theta_{NS}$ measured by $H^3({\bf RP}^4, {\widetilde {\bf Z}})
={\bf Z}_2$ since the fivebrane is a magnetic source of 
the $B_{NS}$-field and thus a transition between 
the $SO(N)$ and the $Sp(N)$ gauge group in boundary CFT.

\subsection{The Baryon Vertex in $SO(N)/Sp(N)$}

The baryon vertex in $SU(N)$ was obtained by wrapping a D4 brane over 
${\bf S}^4$. By analogy, one expects that the baryon vertex in 
$SO(N)$ or $Sp(N)$ will consist of a D4 brane wrapped on ${\bf RP}^4$. 
If we are considering $SO(2k)$ gauge theory, there are $k$ units of 
four-form flux on ${\bf RP}^4$ 
when the D4 brane wraps once on ${\bf RP}^4$. 
But there is no gauge invariant combination, in $SO(2k)$ gauge theory,
of $k$ external quarks to obtain a ``baryon vertex''. 
The baryon vertex of $SO(2k)$ gauge theory should couple 
$2k$ external quarks, not $k$ of them. 

Let $\Phi$ be the map of D4 brane worldvolume $X$ to 
$AdS_7\times{\bf RP}^4$. We must impose the condition that the
$B_{NS}$ fields should be topologically trivial when pulled back to
$X$ as implied in deriving the topological restrictions on the brane
wrapping in section II. 
If one chooses the D4 brane topology as ${\bf RP}^4$, the
condition cannot be satified because of 
$H^3({\bf RP}^4, \widetilde{\Z})={\bf Z}_2$. 
Instead, if we choose the D4 brane topology as ${\bf S}^4$, 
the topological triviality of the field $B_{NS}$ is obeyed 
since $H^3({\bf S}^4, \widetilde{\Z})=0$. Then the map $\Phi:{\bf S}^4
\rightarrow {\bf RP}^4$ gives the degree two map, in other words,   
${\bf S}^4$ wraps twice around ${\bf RP}^4$. Thus we can obtain the
correct baryon vertex coupling $2k$ quarks. 

In $Sp(k)$ gauge theory, a baryon vertex can decay to $k$ mesons. 
Thus, one may expect no topological stability for the $AdS_7$ baryon
vertex. In previous section, we showed that there is a topological 
restriction on the D4 brane wrapping on ${\bf RP}^4$, which is
only possible in $SO(N)$ gauge theory. But, we should consider the
possibility on an existence of nontrivial torsion class of 
the $B_{NS}$ fields due to the topology of the D4 brane 
worldvolume $X$, which is denoted as $W\in H^3(X, \widetilde{\Z})$. 
Then this means that the correct global restriction is not that 
$i^*([H_{NS}])=0$ but rather that
\beq
i^*([H_{NS}])=W,
\eeq
where $i$ is the inclusion of $X$ in spacetime and $[H_{NS}]$ the
characteristic class of the $B_{NS}$ field we have seen in (\ref{gh}). 
A possible $W$ can be determined by using 
the ``connecting homomorphism'' in an exact sequence of cohomology groups 
from the second Stieffel-Whitney class $w_2(X)\in H^2(X, {\Z}_2)$.
If the baryon vertex decays via compact five-manifold $X$, we will find
that $W=0$ since $w_2(X)=0$. Consequently, we cannot obtain the baryon
vertex in the $O4^+$ type under consideration 
due to the global restriction $i^*([H_{NS}])=0$, 
which means that $\theta_{NS}=0$. 

In $SO(2k)$ gauge theory, super Yang-Mills theory with 16 supercharges
actually has $O(2k)$ symmetry, not just $SO(2k)$. The generator $\tau$
of the quotient $O(2k)/SO(2k) = {\bf Z}_2$ behaves as a global
symmetry. Since the baryon is odd under $\tau$, it cannot decay to
mesons which is even under $\tau$. 
Since $H_4({\bf RP}^4, \widetilde{\Z})={\Z}_2$, two baryons can
annihilate into $2k$ mesons. This also comes from the fact that, 
in $O(N)$, a product of two epsilon symbols can be rewritten as a sum
of products of $N$ Kronecker deltas. Another possibility 
is that a baryon vertex constructed from a wrapped D4 brane is
transformed to a state containing a wrapped D2 brane, a ``Pfaffian'' 
state in section IIIA, plus strings
making pairwise connections between external charges since both are
odd under $\tau$. We will show that the latter is not the case. 
 
Consider a configuration of a D2 brane ending on a D4 brane. The end
of the D2 brane on a D4 brane worldvolume is a magnetic source for the
$U(1)$ gauge field $a$ that propagates on the D4 brane. Let $X$ be the
D4 brane worldvolume, $E$ the worldvolume of a D2 brane whose boundary
is on $X$, $D$ the boundary of $E$. Then, the Poincar\'e dual of $D$
is a class $[D]\in H^3(X, \widetilde{\Z})$. Since $D$ acts as a
magnetic source for $a$, the equation which restricts a wrapped 
D4 brane on ${\bf RP}^4$ modifies in the presence of a D2 brane as
follows
\beq
i^*([H_{NS}])=[D].
\eeq
Thus the configuration of the wrapped
D4 brane with a D2 brane ending on it is only possible 
in the $Sp(k)$ gauge theory 
since in this case $\theta_{NS}\neq 0$.

Recall that we need $2k$ elementary strings ending on the wrapped D4
brane in order to cancel $2k$ units of the $U(1)$ charge on the D4
brane and even number of D4 branes, near to infinity, in order to provide
external quarks with even flavors which are necessary for Witten
anomaly \cite{witta} in 5 dimension, $\pi_5(Sp(k))=\Z_2$. 
In $Sp(k)$ gauge theory, when we have taken only two flavors for
simplicity, the baryon itself can decay to $k$ mesons 
\beq
B=\frac{1}{k!}{M_{rs}}^k
\eeq
where $r,\,s=1,2$ are flavor indices and 
$M_{rs}=\frac{1}{2}\gamma_{ab}\psi^a_r\psi^b_s$ is a meson 
formed by the invariant second rank antisymmetric tensor $\gamma^{ab}$.
Thus it is expected that the natural decay channel
of the baryon vertex constructed from the above configuration 
will be a state containing $k$ mesons where $2k$ charges on the
boundary are connected pairwise by elementary strings and 
the wrapped D2 brane on ${\bf RP}^2 \subset {\bf RP}^4$: 
\beq
\frac{1}{k!}{M_{rs}}^k
\oplus \frac{1}{k}\mbox{Tr}\Phi^{i_1}\cdots
\Phi^{i_k},
\eeq
where $i$'s are indices of $R$-symmetry group.

\section{Discussion }
\setcounter{equation}{0}

To summarize, for $SU(N)$ $(0, 2)$ theory,
we interpreted the baryon vertex as a wrapped D4 brane in $\S^4$ 
connected by $N$ fundamental strings ending on a D4 brane on $AdS_7$ 
boundary.
When we go $SO(2N)$ $(0, 2)$ theory, $\R^5/\Z_2$ orbifold singularity was 
crucial to understand M theory realization of four types of O4 plane.
We constructed the possible brane wrappings on $\RP^4$ in Type IIA string 
theory and determined their topological restrictions in each case.  
According to this classification, it was possible to interpret 
various wrapping branes on $\RP^4$
in terms of 
particles, strings, twobranes (in $AdS_7$), domain walls and the baryon 
vertex where the topological properties on $\RP^4$ are heavily used.

When we go further compactification of $(0,2)$ theory on a circle 
which would be interpreted as the compactified time parameter 
(in Euclidean space), 
so temperature, then the low energy theory will be four
dimensional ${\cal N}=4$ super Yang-Mills theory. 
If we instead insist on the boundary condition breaking the
supersymmery by taking the fermions to be antiperiodic in going around
the circle, the low energy theory will be the pure Yang-Mills theory
without supersymmetry. Witten proposed \cite{wittqcd4} 
that this $QCD_4$ theory has
the dual description by the AdS Schwarzschild soution. 
According to the conjecture, the compactified six dimensional $(0,2)$
theory considered in this paper can be reduced to the $QCD_4$ theory 
by introducing a supersymmetry breaking circle. 
Thus the various spectrums of the 5 dimensional super Yang-Mills
theory should be related to 4 dimensional QCD spectrums.
It will be then interesting to study $QCD_4$ along this line. 

Recently, Aharony and Witten \cite{aw} showed that, 
if $SU(N)\;(0,2)$ theory is
compactified on two torus ${\bf T}^2$ or any Riemann surface, 
the theory has a ${\Z}_N$ global symmetry and this provides a test of
the AdS/CFT correspondence for finite $N$.
It is also interesting to
study whether our $SO(2N)$ $(0, 2)$ theory compactified on $\S^1 \times
\S^1$ should have also such topological symmetry, in which case
the $SO(2N)$ gauge group has a ${\Z}_2$ center.

It is natural to ask what happens for the other well known 
M theory on $\R^3 \times (\R^7\times \S^1)/\Z_2$,
where $\S^1$ is the eleventh dimensional cirlce.
In this case, M2 branes are put at one of orbifold singularities,
i.e. the origin of $\R^7$ and $\S^1$ coordinates.
In Type IIA viewpoint, the above configuration will appear as D2
branes on the $\R^7/\Z_2$ orientifold singularity, which produces no
orientifold 2-plane when the size of $\S^1$ goes to infinity.
%\footnote{We thank O. Aharony for pointing out this.} 
Then we expect to have the 3 dimensional $SO(2N)$ maximally
supersymmetric SYM theory, which flows in the IR to the
${\cal N}=8$ SCFT corresponding to the M2 branes on the orbifold point
$\R^8/\Z_2$ \cite{aoy}.
Along the same strategy as the case $AdS_7 \times \RP^4$,
one can also analyze various brane configurations
in Type IIA string theory.
It would be interesting to elaborate this further in the future.

\vspace{1cm}

\centerline{\bf Acknowledgments} 

We thank O. Aharony and Y. Oz for email correspondence.
HSY thanks Prof. B.-H. Lee for helpful discussions.
This work of CA and HSY is supported (in part) by the Korea Science 
and Engineering Foundation (KOSEF) through the Center 
for Theoretical Physics (CTP) at Seoul National University and 
HSY is also partially supported by the Korean Ministry of Education 
(BSRI-98-2414) and HK is supported by TGRC-KOSEF. 
We thank Asia Pacific Center for Theoretical 
Physics (APCTP) for partial support and hospitality 
where this work has been done. 

%****************************************************
%****************************************************
%****************************************************

\end{document}